          [2004/06/22 1.2 (KOF); 2001/04/25 1.1 (PWD)]

\documentclass[a4paper,twocolumn]{Gaia2004} 
\usepackage{times}      
\usepackage{epsfig}     
\usepackage{natbib}     
\title{Extracting stellar parameters from observed spectra: 
the role of cross-correlation and minimum distance methods}

\author{Toma\v{z} Zwitter}
\affil{University of Ljubljana, Dept. of Physics, Ljubljana, Slovenia}
\author{Ulisse Munari}
\affil{National Institute of Astronomy, Padua Astronomical Observatory 
       in Asiago, Italy}
\author{Arnaud Siebert}
\affil{Steward Observatory, Tucson, AZ, USA}

\bibpunct{(}{)}{;}{a}{}{,}  

\begin{document}


\maketitle

\begin{abstract}
One may consider cross-correlation routines applied to stellar spectra 
as a way to determine not only the radial velocities but also to  assess the  
properties of stellar atmospheres. We show that this is not the case. 
High value of the cross-correlation coefficient between the observed and 
a template spectrum does not mean that the template corresponds to correct 
values of stellar temperature, gravity or metallicity. Cross-correlation 
operates in normalized space, so it is not sensitive to scaling or shifting 
of stellar flux. It is argued that cross-correlation is useful to efficiently
determine the velocity shift between the observed and the template spectrum. 
But minimal distance methods need to be used to determine which template 
gives the best match to the observed spectrum, and so derive values of 
physical parameters. To illustrate the point we use data from the 
RAVE project which are being obtained at very similar spectral 
resolution and wavelength domain as will be the case for GAIA's RVS.
\end{abstract}

\section{Introduction}

In the attempt to extract values of 
physical parameters from stellar spectra one first tries to 
determine the value of radial velocity. If the spectra have a wide wavelength 
coverage, like in the case of ELODIE spectra, it can be assumed that a fixed 
spectral mask gives radial velocities with sufficient accuracy (Katz et al. 
1998). Analysis of GAIA radial velocity spectrometer (RVS) data will profit from 
quite accurate knowledge of stellar parameters obtained from on-board 
photometric observations. So all simulations of GAIA radial velocity 
determination assumed that the choice of a proper template is 
given by complimentary photometric data (Katz et al.\ 2004, 
Munari et al. 2003, Zwitter 2002). Alternatively it was assumed, while assessing
the capabilities of GAIA spectra to independently determine the values of stellar 
parameters, that the radial velocity is a-priori known 
(ESA-2000-SCI-4, Thevenin et al. 2003, Katz et al. 2004). Note that these 
approaches are relevant at the end of the GAIA mission, 
but at the start of the mission virtually nothing will be known about the values 
of stellar parameters, so that one would not know which kind of template to 
use for a radial velocity determination. Also, the wavelength coverage of 
GAIA spectra is much smaller than that of a typical Echelle (ELODIE) type 
spectrum. So we attempt here to determine {\it both\/} the best radial 
velocity and the values of parameters of the stellar atmosphere simultaneously.
Data from the RAVE experiment, which are very similar in both wavelength 
coverage and resolution to those of GAIA's RVS, will be used to illustrate the point. 
Note that since virtually nothing is known 
about the RAVE targets it is obvious to use such an approach. 

\section{Observed spectra}

RAVE (RAdial Velocity Experiment) project (Steinmetz 2003, Munari et
al. 2004) is an ambitious international collaboration aimed
at conducting an all-sky spectroscopic survey of Galactic stars. 
The participating countries are Australia, Canada, France, Germany, 
Italy, Japan, Netherlands, Slovenia, Switzerland, UK, and USA. The
spectra allow us to measure radial velocities as well as metallicites and
other stellar parameters for a large number of stars using the 1.2-m UK
Schmidt telescope at the Anglo-Australian Observatory.  Data collection
for this project started in April 2003 with more than 40,000 spectra
observed to date. Spectral resolving power and 
wavelength range (8413--8746~\AA) are very similar to those of 
the GAIA Radial Velocity Spectrograph.

\begin{figure*}[!ht]
\begin{center}
\leavevmode
\centerline{\epsfig{file=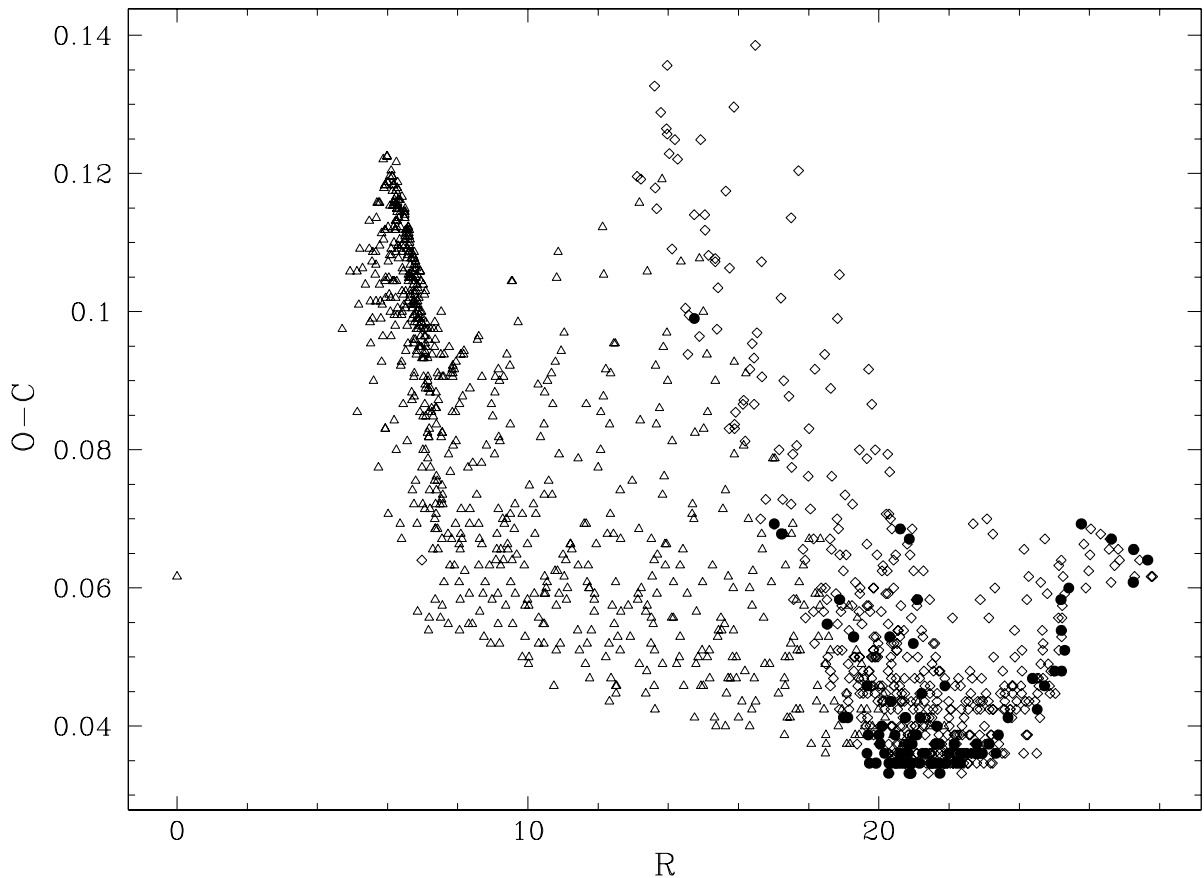,angle=270,width=\textwidth}}
\end{center}
\caption{Fits of templates with the best value of correlation 
coefficient $R$ are not necessarily those with the smallest 
value of the standard deviation (O-C) between the observed and 
template spectra. The later is given in units of continuum. 
Each point presents results of a cross-correlation and standard deviation 
for one template from the database. Symbol type marks the difference between 
the calculated value of radial velocity for a given template and the 
one for a minimal O-C template: less than 0.2~km~s$^{-1}$ ($\bullet $), up to 
2~km~s$^{-1}$ ($\diamond$), or more ($\triangle$).
}
\label{f1}
\end{figure*}

\begin{figure*}[!ht]
\begin{center}
\leavevmode
\centerline{\epsfig{file=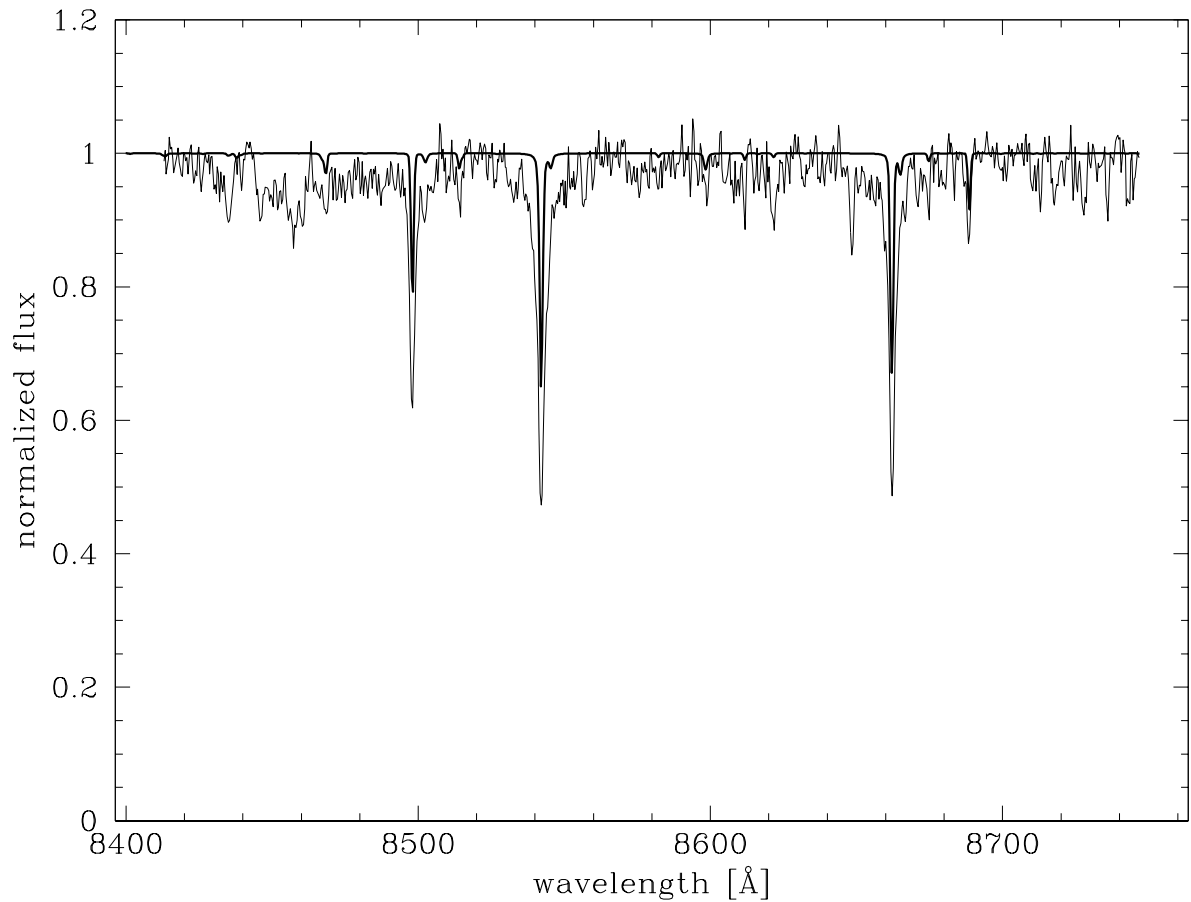,angle=270,width=\textwidth}}
\end{center}
\caption{Match between the observed (thin-line) and template spectrum 
(thick-line) is poor, even though this is the template which yielded the 
highest value of the correlation coefficient $R$ in Figure~1.}
\label{f1}
\end{figure*}

\begin{figure*}[!ht]
\begin{center}
\leavevmode
\centerline{\epsfig{file=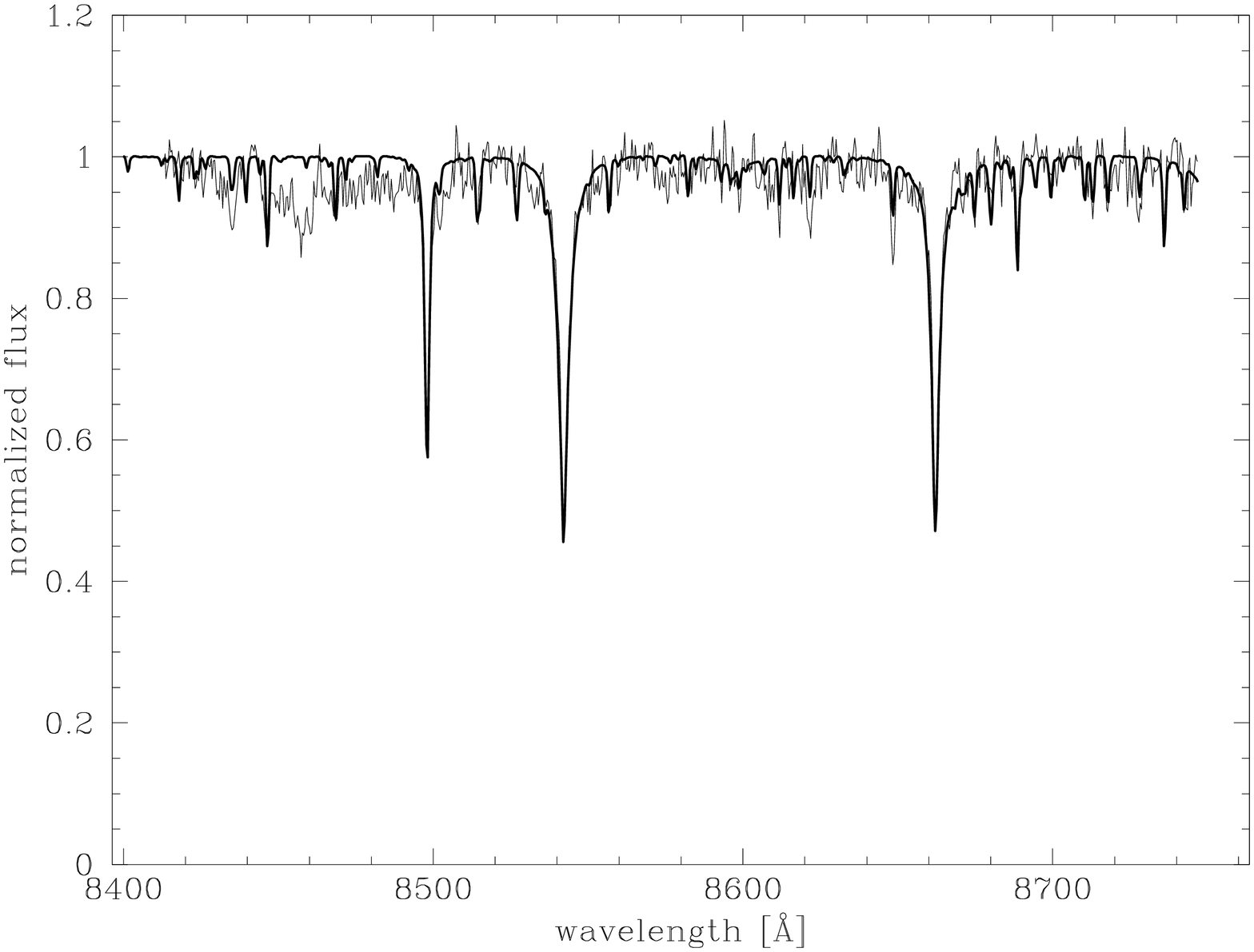,angle=270,width=\textwidth}}
\end{center}
\caption{
The match between the observed (thin-line) 
and template spectrum (thick-line) is much better 
when a suitable minimization scheme is used as a figure of merit. 
The template and observed spectrum correspond to the lowest $O-C$ point
in Figure~1.
}
\label{f1}
\end{figure*}

\section{Library of synthetic spectra}

Observed or synthetic spectra can be used for classification. The advantage of 
the former is that they are "real", so avoiding any simplifying assumptions 
used in synthetic spectra calculation (Munari et al. 2001). The latter have the advantage 
of a more uniform coverage of the parameter space. At present we are using a subset of a 
synthetic library of nearly 60.000 spectra with suitable resolving power 
(Zwitter, Castelli \&\ Munari 2004) 
which were calculated from the latest generation of Kurucz models. The grid is characterized 
by the following ranges of stellar parameters: 
$3500 \, \mathrm{K} \le \mathrm{T_{eff}} \le 47500 \, \mathrm{K}$, 
$0.0 \le \log g \le 5.0$, 
$-3.0 \le [\mathrm{M}/\mathrm{H}] \le +0.5$, 
$0 \le \mathrm{V_{rot}} \le 500 $~km~s$^{-1}$, 
$\xi = 2 $~km~s$^{-1}$.

\section{Using correlation routines only}

Correlation routines, like the IRAF's rvsao package (Kurtz \&\ Mink 1998) are a 
well established way to calculate radial velocity using a fixed stellar or 
galaxy spectrum template. Their figure of merit is based on the value of 
correlation coefficient $R$ (Tonry \&\ Davis 1979). But one should not conclude 
that the template spectrum with the highest value of $R$  also gives the best fit 
to the observed spectrum. Fig.~1 shows results of analysis of a single 
observed spectrum using a set of template spectra. The maximal value of the 
correlation coefficient $R$ occurs at the ``stem'' of area covering a 
caravel ship shape, though the minimum of $O-C$ occurs at its bottom. 
The reason is that the correlation is defined using normalized quantities. 
Therefore it is not sensitive to any kind of constant shift or spread 
(of the kind $x+a$ and $x*a$ where $a$ is a constant 
value or bias). This explains why it is only sensitive to high variations in 
the line profile and is not suitable for minimization. 

We can also say that in cross-correlation one tries to find the best alignment 
of the lines, in a way irrespective of their depth. This is strictly valid for 
isolated single lines. The effective wavelength of a blend of nearby lines can 
shift sideways and so worsens the cross-correlation only with a large change 
of the temperature and gravity. 

Figure 2 plots the observed spectrum which was used in Figure 1 and one 
of the templates from the library. The match is poor even though the 
template shown features the 
highest value of the correlation coefficient $R$ among all templates in the 
library (the point is at the tip of the stem of the caravel shape in Figure 1). 
Cross-correlation alone is 
clearly not enough to find the template with the best match.

\section{Adding a minimal distance method}

The situation can be improved by supplementing correlation analysis with a 
proper minimization scheme (Figure 3). The reason is that 
normalized squared distance is sensitive to line depth and shape as well 
as continuum level, and so to the values of stellar parameters. 

In our case we first use a correlation routine to calculate the velocity 
shift between the observed spectrum and a template from 
the synthetic library. But the goodness of fit is not based on the value 
of the correlation coefficient R. Instead we calculate the standard 
deviation between the observed spectrum and the velocity-shifted template. 
The template which yields the smallest standard deviation is assumed to give 
a fair representation of parameters of the stellar atmosphere. 

\section{Relevance of RAVE, GAIA and other automated classification 
projects}

Values of several stellar parameters are primarily based on spectral 
analysis. These include metallicity, abundances of individual elements 
and stellar rotation velocity. The values of effective temperature and 
surface gravity can also be derived, or in the case of GAIA checked, by 
spectral analysis. Both GAIA and RAVE projects will provide too many 
spectra to be analyzed by hand. Automated procedures, such as the one 
proposed here, yield useful results. Any other information available, 
in the case of GAIA these would be results of photometric and astrometric 
observations, can be seen as an additional constraint on the relevant 
range of stellar templates to be considered for a given star. 
If an acceptable match could not be 
found despite a high S/N of the observed spectrum, this implies that any kind 
of peculiarity is present in the observed object, or that the observed star 
is a double lined spectroscopic binary.

\section{Conclusion}

Cross correlation is sensitive to position of the lines and so radial 
velocity. It is however not sensitive to line depth, so it is not to be 
used to determine the values of stellar atmosphere parameters unless 
blends of nearby lines are driving the correlation. The situation gets 
worse with low S/N spectra and when sharp lines (e.g. Ca II IR triplet) 
are present. By adding a suitable minimum distance scheme which is sensitive 
to equivalent widths (Thevenin \&\ Foy 1983) the values of both radial 
velocity and stellar parameters like temperature, gravity, metallicity 
and rotation velocity can be determined reliably. 

\section*{Acknowledgments}
TZ acknowledges financial support from the Slovenian Ministry 
for Education, Science and Sports.


\begin{thebibliography}{}
\bibitem [\protect\astroncite 
{Katz et~al.}{1998}]{Katzetal1998}
Katz, D., Soubiran, C., Cayrel, R., Adda, M., Cautain, R. 1998, A\&A 338, 151 
\bibitem [\protect\astroncite 
{Katz et~al.}{2004}]{Katzetal2004}
Katz, D., et~al. 2004, MNRAS, in press 
\bibitem [\protect\astroncite 
{Kurtz \& Mink}{1998}]{KurtzMinkl1998}
Kurtz, M.J., Mink, D.J. 1998, PASP 110, 934
\bibitem [\protect\astroncite 
{Munari et~al.}{2001}]{Munarietal2001}
Munari, U., Agnolin, P., Tomasella, L. 2001, Balt.A. 10, 613
\bibitem [\protect\astroncite 
{Munari et~al.}{2003}]{Munarietal2003}
Munari, U., Zwitter, T., Katz, D., Cropper, M. 2003, In GAIA Spectroscopy: 
  Science and Technology, ASP Conf. Ser. 298, 275 
\bibitem [\protect\astroncite 
{Munari et~al.}{2004}]{Munarietal2004}
Munari, U., Zwitter, T., Siebert, A. 2004, this volume
\bibitem [\protect\astroncite 
{Steinmetz}{2003}]{Steinmetz2003}
Steinmetz, M., 2003, In GAIA Spectroscopy: Science and 
Technology, ASP Conf. Ser. 298, 381 
\bibitem [\protect\astroncite 
{Thevenin \& Foy}{1983}]{Thevenin1983}
Thevenin, F., Foy, R. 1983, A\&A 122, 261
\bibitem [\protect\astroncite 
{Thevenin et~al.}{2003}]{Theveninetal2003}
Thevenin, F., Bijaoui, A., Katz, D. 2003, In GAIA Spectroscopy: Science and 
Technology, ASP Conf. Ser. 298, 291 
\bibitem [\protect\astroncite 
{Tonry \& Davis}{1979}]{Tonry1979}
Tonry, J., Davis, M. 1979, AJ 84, 1511 
\bibitem [\protect\astroncite 
{Zwitter}{2002}]{Zwitter2002}
Zwitter, T. 2002, A\&A 386, 748
\bibitem [\protect\astroncite 
{Zwitter et~al.}{2004}]{Zwitteretal2004}
Zwitter, T., Castelli, F., Munari, U. 2004, A\&A, 417, 1055 
 \end{thebibliography}
\end{document}